\documentclass[a4paper,11pt]{article}
\usepackage{pos}

\title{CP, or not CP, that is the question \ldots}

\author*[a]{Andreas Ringwald}

\affiliation[a]{Deutsches Elektronen-Synchrotron DESY,\\
  Notkestr. 85, 22607 Hamburg, Germany}

\emailAdd{andreas.ringwald@desy.de}

\abstract{
 Motivated by recent claims questioning the existence of strong CP violation, we present a pedagogical review of CP violation in Quantum Chromodynamics (QCD). Using fundamental properties of the QCD partition function, we analyze the dependence of the chiral quark and CP violating gluon condensates on the $\theta$ parameter and the quark masses in the chiral limit. We show explicitly how CP violation arises, clarify the role of the axial U(1) anomaly and the ordering of the infinite-volume limit, and discuss the conditions under which CP symmetry may or may not be realized, including in the large-$N_c$ framework. Our results reaffirm the presence of strong CP violation for physically relevant parameters and thus the theoretical basis of the strong CP problem and axion physics.
}

\FullConference{3rd General Meeting of the COST Action COSMIC WISPers (CA21106)  (COSMICWISPers2025) and\\
                3rd Training School of the COST Action COSMIC WISPers (CA21106) (COSMICWISPers2025)
\\
9–12 Sept 2025 and 16–19 Sept 2025\\
Sofia, Bulgaria and Annecy, France\\}

\begin{document}
\maketitle


Our COST Action COSMIC WISPers focusses 
on an exhaustive study of very weakly interacting slim particles (WISPs). 
Among those WISPs, the axion is particularly well motivated, since it 
was originally proposed to solve the strong CP puzzle of Quantum Chromodynamics (QCD), namely 
why CP violation -- although allowed by the parameters in 
the QCD Lagrangian  
-- is experimentally constrained to be extremely small~\cite{Baluni:1978rf,Crewther:1979pi,Abel:2020pzs}. The Peccei–Quinn mechanism
dynamically explains this smallness, and the axion emerges as its unavoidable consequence.
Unlike many other WISPs, the axion is therefore not introduced ad hoc, but is motivated by a well-defined theoretical puzzle. 

Recently, the very claim that QCD possibly features CP violation was challenged~\cite{Ai:2020ptm,Ai:2024cnp}. It is the purpose of this proceedings contribution to present a pedagogical review of the calculation of CP violation in QCD. We perform it in a way which allows to compare with the recent claim of the absence of strong CP violation 
in QCD.

The Lagrangian of QCD is the one of an SU(3) Yang-Mills theory with colored quarks,
\begin{eqnarray}
{\cal L} = 
-\frac{1}{4}  G_{\mu\nu}^a G^{\mu\nu}_a 
+ \bar{q}\, i \gamma_\mu \left(\partial^\mu -i g G^\mu \right) q  
-\bar{q}_{\rm R} {\mathcal M}\,
q_{\rm L} 
-\bar{q}_{\rm L}{\mathcal M}^\dagger
q_{\rm R} 
-  \theta \frac{g^2}{32\pi^2}  G_{\mu\nu}^a\tilde{G}^{\mu\nu}_a,
\end{eqnarray}
given in terms of gluon field strengths $G_{\mu\nu}^a$ and their duals $\tilde G_{\mu\nu}^a \equiv \frac{1}{2} \epsilon_{\mu\nu\rho\sigma} G^{\rho\sigma}_a$, with $a = 1,..., N_c^2-1$, for $N_c = 3$, 
quark spinors  $q$  (flavour indices  $u$, $d$, $s$, \ldots suppressed) and their right- and left-handed projections, $q_{\rm R} \equiv \frac{1}{2} \left( 1 + \gamma_5\right) q$     
and $q_{\rm L} \equiv \frac{1}{2} \left( 1 - \gamma_5\right) q$, respectively, and the 
following parameters: 
{\em i)}
the intrinsic scale $\Lambda$, related to the strong coupling $g^2(Q) \to  \frac{8\pi^2}{\beta_0 \ln (\Lambda/Q) }$, for $Q\gg \Lambda$, with $\beta_0 = \frac{11}{3} N_c - \frac{2}{3} N_f$, 
{\em ii)}
the $N_f\times N_f$ quark mass matrix $\mathcal M$, which is in general non-diagonal and non-Hermitian, $\mathcal M \neq \mathcal M^\dagger$, and whose form depends on the choice of the quark field basis, and 
{\em iii)} the theta parameter $\theta$. 

On the classical level, if the masses of the three light (masses $m_i<\Lambda$) quark flavors $u$, $d$, and $s$ are neglected, QCD has a $\rm U(3)_R \times U(3)_L = SU(3)_R \times SU(3)_L \times U(1)_{R+L} \times U(1)_{R-L}$          symmetry under independent rotations of right- and left-handed quark flavors, $q_{\rm R}^\prime = V_{\rm R} q_{\rm R},\ q_{\rm L}^\prime = V_{\rm L} q_{\rm L}$. 
On the quantum level, the conservation law for the singlet axial current associated to           $U(1)_{\rm R-L}$ contains an anomaly:
\begin{equation}
 \partial_\mu \left(  \bar{q} \,\gamma^\mu \gamma_5 \, q \right) = 2 \bar q\, {\mathcal M}\, i\gamma_5 q + 2 N_f \frac{g^2}{32\pi^2} G_{\mu\nu}^a \tilde G^{\mu\nu}_a .
 \end{equation}
 Hence, the symmetry group of QCD with 3 massless flavors is $\rm SU(3)_R\times SU(3)_L\times U(1)_{R+L}$.
This is why QCD with 3 massless flavors has only  $N_f^2 -1 = 8$ pseudo-scalar Goldstone bosons from the breaking of  $\rm SU(3)_R\times SU(3)_L\times U(1)_{R+L}$                                              to $\rm SU(3)_{R+L}\times U(1)_{R+L}$. 
There is no pseudo-scalar Goldstone boson from the breaking of $\rm U(1)_{R-L}$,  
because the latter symmetry is anomalous.    

Chiral transformations generated by the factor $\rm U(1)_{R-L}$ 
change the phase of right-handed components of all quark fields by same angle, $q_{\rm R}^\prime = e^{i\alpha} q_{\rm R}$,  while the left-handed components are subject to the opposite transformation: $q_{\rm L}^\prime = e^{-i\alpha} q_{\rm L}$. 
This change of basis can be compensated by modifying the quark mass matrix with ${\mathcal M}^\prime =  {\mathcal M} e^{2i\alpha}$, but in view of the anomaly, the operation does not represent a symmetry of the system.
If the above change of the quark mass matrix is accompanied by a simultaneous change of the $\theta$ parameter to   $\theta^\prime = \theta - 2\alpha$, the physics does remain the same. 
Starting from an arbitrary mass matrix, a change of basis involving the factor $\rm U(1)_{R-L}$                is needed to arrive at the convention where ${\mathcal M}$      is diagonal with real eigenvalues. In that convention, the theta parameter does have physical significance – otherwise only the product  $ {\mathcal M}e^{i\theta/N_f}$ counts.

The investigation whether QCD possibly violates CP  has to rely on non-perturbative methods. 
Here, we rely on a non-perturbative evaluation of the partition function of QCD -- the thermal trace  
\begin{eqnarray}
    Z = 
    {\rm Tr} \left\{\exp \left( -\beta H \right) \right\},
\end{eqnarray}
where $\beta \equiv 1/T$ is the inverse temperature and $H$ is the QCD Hamiltonian. Since we
are interested in zero-temperature QCD, we will take the limit $\beta \to \infty$
at the end.  
The partition function can be expressed as a path integral in Euclidean space, 
\begin{eqnarray}
    Z  &=& 
    \int [ {\rm d}G ] [ {\rm d}\bar q ][ {\rm d}q ]
\exp \left[ - \int {\rm d}^4 x \left\{ {\cal L}^0
+\bar{q}_{\mbox{\scriptsize R}} {\mathcal M}\,
q_{\mbox{\scriptsize L}} 
+\bar{q}_{\mbox{\scriptsize L}}{\mathcal M}^\dagger 
q_{\mbox{\scriptsize R}}  
- i 
\theta  \frac{g^2}{32\pi^2}\, G_{\mu\nu}^a\tilde{G}_{\mu\nu}^a 
\right\}
\right],
\end{eqnarray}
\begin{eqnarray}
{\cal L}^0 &=&
 \frac{1}{4} \, G_{\mu\nu}^a G_{\mu\nu}^a -  i \bar{q} \gamma_\mu (\partial_\mu -igG_\mu )q  .
\end{eqnarray}
In the time-direction, $x_4$, the gluon fields are required to be periodic up to a gauge transformation, whereas the quark fields have to obey antiperiodic boundary conditions. It is convenient to impose the same boundary conditions also in the three spatial directions, so that the fields are effectively defined on a four-dimensional torus of size $L_1\times L_2\times L_3\times \beta$ and four-volume $V=L_1 L_2 L_3\, \beta$.

From the path integral representation it is obvious that the ``free energy'' $F \equiv - \ln Z$ associated with the partition function
plays the role of a generating functional of the connected correlation functions of the chiral quark densities $\bar q_{\rm R}q_{\rm L}$, $\bar q_{\rm L}q_{\rm R}$, and the pseudo-scalar gluon density $\frac{g^2}{32\pi^2} G_{\mu\nu}^a \tilde G_{\mu\nu}^a$. In particular, 
the chiral quark condensates and the pseudo-scalar gluon condensate are obtained from first  functional derivatives of $F$: 
\begin{eqnarray}
\label{eq:chiral_quark_condensates}
      \frac{\delta F}{\delta \mathcal M} = - \frac{1}{Z}\frac{\delta Z}{\delta \mathcal M} 
     = \langle \bar q_{\rm R} q_{\rm L}\rangle ,\hspace{3ex} 
   \frac{\delta F}{\delta \mathcal M^\dagger} = - \frac{1}{Z}\frac{\delta Z}{\delta \mathcal M^\dagger} = \langle \bar q_{\rm L} q_{\rm R}\rangle ,
 \end{eqnarray}
 \begin{eqnarray}
 \label{eq:pseudo_scalar_gluon_condensate}
  i\frac{\delta F}{\delta \theta} = - i\frac{1}{Z}\frac{\delta Z}{\delta \mathcal \theta}
= \left\langle \frac{g^2}{32\pi^2} G_{\mu\nu}^a \tilde G_{\mu\nu}^a \right\rangle 
.
\end{eqnarray}
For $V\to \infty$, these condensates approach vacuum expectation values. 
A non-zero value of the pseudo-scalar gluon condensate signals P- and T-, and thus CP-violation 
in strong interactions~\cite{Shifman:1979if}.
Further progress can be made by evaluating the free energy $F$ further in the chiral limit, ${\cal M}\to 0$. The methods to perform this task have been exposed in a seminal paper by  Leutwyler and Smilga in  Ref.~\cite{Leutwyler:1992yt}, on which we rely heavily in the following. 

Let us start the discussion with QCD with $N_f=1$ quark flavor of mass $m$. 
This theory is expected to have a mass gap $M_0 \sim \Lambda$ which  
persists in the chiral limit, $m\to 0$. 
In fact, as discussed before, the $\rm U(1)_{R-L}$ symmetry is broken not spontaneously but explicitly by the anomaly and therefore no corresponding Goldstone-boson is expected to appear.
The mass gap implies a finite correlation length $\xi\sim M_0^{-1}$. Therefore, the free energy $F$ is an extensive quantity, and the   
free energy density $f\equiv F/V$ tends, for $V  \gg M_0^{-4}\sim \Lambda^{-4}$, to the vacuum energy density, $\epsilon_{\rm vac}$:\footnote{The finite volume corrections, $f (V)- \epsilon_{\rm vac}$, are exponentially suppressed, 
$\propto \exp (- M_0 L)$.}
\begin{eqnarray}
\label{eq:free_energy_density_one_flavor}
 f = \frac{F}{V} = - \frac{\ln Z}{V}  =    \epsilon_{\rm vac} \left( me^{i\theta} \right)  
,\hspace{3ex} {\rm for\hspace{3ex} }\ V  \gg\Lambda^{-4} .
\end{eqnarray}
Importantly, as indicated here, the vacuum energy density inherits from the partition function that it can depend on the quark mass $m$  
and the $\theta$ parameter only through the product $m e^{i\theta}\equiv z$. 
Moreover, the persistence of the mass gap in the chiral limit, together with the fact 
that the vacuum energy is real, implies that $\epsilon_{\rm vac}(z)$ is real-analytic 
around $z=0$, that is it has a Taylor expansion 
$\epsilon_{\rm vac}(z)=a_0 + a_1 (z+z^\ast )+ \ldots$, with real coefficients $a_i$, around $z=0$. 
Correspondingly, up to linear order in $m$, the vacuum energy density can be parametrised as~\cite{Leutwyler:1992yt},
\begin{eqnarray}
\label{eq:vac_energy_one_flavor_chiral_expansion}
     \epsilon_{\rm vac} \left(m e^{i\theta}\right) = \epsilon_{\rm vac}(0) - \Sigma\  \Re \left( m e^{i\theta} \right) + {\mathcal O}(m^2), 
\end{eqnarray}
where the zeroth order term $\epsilon_{\rm vac}(0)$ just affects the overall normalisaton of the partition function. 
The real low-energy constant $\Sigma \sim \Lambda^3$ is related to the quark condensate in the chiral limit. In fact, one can infer from  Eqs.~\eqref{eq:chiral_quark_condensates}, \eqref{eq:free_energy_density_one_flavor}, 
and \eqref{eq:vac_energy_one_flavor_chiral_expansion}, that\footnote{These expressions corroborate the findings of Crewther~\cite{Crewther:1979rs}  
who evaluated directly the chiral condensate in the chiral limit by exploiting the anomalous Ward identities of the $\rm U(1)_{R-L}$ current and the absence of a $\rm U(1)_{R-L}$ 
boson.} 
\begin{eqnarray}
\label{eq:chiral_quark_condensates_one_flavor_rl}
\langle \bar q_{\rm R} q_{\rm L}\rangle_{|m=0} &=& 
- 
\frac{\partial}{\partial m}\frac{\ln Z}{V}
= 
\frac{\partial}{\partial m} \left\{\Sigma \Re (m e^{i\theta} )+ \mathcal{O}(m^2)\right\}
= 
\frac{1}{2}
\Sigma e^{i\theta} + \mathcal{O}(m),
\\
\label{eq:chiral_quark_condensates_one_flavor_lr}
\langle \bar q_{\rm L} q_{\rm R}\rangle_{|m=0} &=& 
-  
\frac{\partial}{\partial m^\ast}\frac{\ln Z}{V}
= 
\frac{\partial}{\partial m^\ast} \left\{\Sigma \Re (m e^{i\theta} )+ \mathcal{O}(m^2)\right\}
= 
-\frac{1}{2}
\Sigma e^{-i\theta} + \mathcal{O}(m),
\end{eqnarray}
corresponding to
\begin{equation}
\label{eq:quark_condensates_one_flavor}
    \langle \bar q q\rangle = - \Sigma \cos\theta + \mathcal{O}(m), \hspace{6ex}
\langle \bar q\, i\gamma_5 q\rangle = - \Sigma \sin\theta + \mathcal{O}(m)
.
\end{equation}  
Moreover, and most importantly, the condensate of the pseudo-scalar gluon density is predicted, 
from Eqs.~\eqref{eq:pseudo_scalar_gluon_condensate}, \eqref{eq:free_energy_density_one_flavor}, 
and \eqref{eq:vac_energy_one_flavor_chiral_expansion} as 
\begin{equation}
\label{eq:pseudo_scalar_gluon_condensate_one_flavor}
    \left\langle  \frac{g^2}{32\pi^2} G_{\mu\nu}^a \tilde G_{\mu\nu}^a \right\rangle = 
-i \frac{1}{V} 
\frac{\partial}{\partial \theta}\ln Z
= 
i\frac{\partial}{\partial \theta} \left\{\Sigma \Re (m e^{i\theta} )+ \mathcal{O}(m^2)\right\}
= \Sigma\, \Im \left( m e^{i\theta}  \right) + \mathcal{O}(m^2) .
\end{equation}
We conclude that $N_f=1$ QCD predicts CP violation as long as 
\begin{eqnarray}
    \Sigma \,\Im \left( m e^{i\theta}  \right) = \Sigma |m| \Im \left( e^{i (\arg(m) + \theta )}  \right) = \Sigma |m| \sin (\underbrace{\theta + \arg(m)}_{\bar\theta}) = \Sigma |m| \sin \bar\theta \neq 0 . 
\end{eqnarray}
In other words: $N_f=1$ QCD violates CP if simultaneously {\em i)} chiral symmetry is broken in the chiral limit ($\Sigma\neq 0$), {\em ii)} the quark mass is non-zero ($|m|\neq 0$), and {\em iii)} the angle $\bar\theta \equiv \theta + \arg (m)$ is not a multiple of $\pi$ ($\sin\bar\theta\neq 0$).

Note that this finding follows directly from fundamental properties of the partition function of 
$N_f=1$~QCD, in particular the extensitivity and analyticity of its corresponding free energy, following in turn from the persistence of the mass gap in the chiral limit and the parameter dependence implications from the $\rm U(1)_{R-L}$ anomaly. One had not to use the fact 
that the path integral corresponding to the partition function can be cast in the form 
of a Fourier-decomposition,
\begin{equation}
    Z = \sum_{\nu = -\infty}^{\infty}  e^{i\nu \theta} Z_\nu ,
\end{equation} 
in terms of path-integrals 
\begin{equation}
    Z_\nu  = 
    \int_\nu [ {\rm d}G ] [ {\rm d}\bar q ][ {\rm d}q ]
\exp \left[ - \int {\rm d}^4 x \left\{ {\cal L}^0
+\bar{q}_{\mbox{\scriptsize R}} {\mathcal M}\,
q_{\mbox{\scriptsize L}} 
+\bar{q}_{\mbox{\scriptsize L}}{\mathcal M}^\dagger 
q_{\mbox{\scriptsize R}}  
\right\}
\right] 
\end{equation}
over gauge fields with fixed winding number\footnote{Gauge fields on a four-torus may be represented by gauge potentials $G_\mu$ defined on the full Euclidean space, supplemented by transition functions $\Omega(x)$ that encode how the fields are related when moving between adjacent periodic cells. These transition functions can be non-trivial, since the gauge potentials $G_\mu$ are required to be periodic only up to gauge transformations. Non-trivial winding numbers arise precisely from such non-trivial transition functions~\cite{tHooft:1981nnx}.} $\nu \equiv \frac{g^2}{32\pi^2} \int_V d^4x \,G_{\mu\nu}^a\tilde{G}_{\mu\nu}^a \in \mathbb{Z}$.
However, one can turn the tables and determine the contribution of the sector with winding number $\nu$ to the partition function by calculating the corresponding Fourier coefficients, 
\begin{equation}
    Z_\nu = \frac{1}{2\pi} \int_{-\pi}^\pi {\rm d\theta} e^{-i\nu \theta} Z(me^{i\theta} ), 
\end{equation}
in terms of  
the explicit expression, 
\begin{equation}
\label{eq:Z_one_flavor}
Z(me^{i\theta}) = \exp \left\{ V \Sigma\  \Re \left( m e^{i\theta} \right) \right\}
,
\hspace{3ex} {\rm for\hspace{3ex} }\ V  \gg \Lambda^{-4} \ {\rm and\ } |m|\ll \Lambda ,
\end{equation}
for the partition function from Eqs.~\eqref{eq:free_energy_density_one_flavor} and \eqref{eq:vac_energy_one_flavor_chiral_expansion}. One finds~\cite{Leutwyler:1992yt}        
\begin{equation}
\label{eq:Z_nu_one_flavor}
    Z_\nu = \left( \frac{m}{|m|} \right)^\nu I_\nu ( V \Sigma |m| ) , 
    \hspace{3ex} {\rm for\hspace{3ex} }\ V  \gg \Lambda^{-4} \ {\rm and\ } |m|\ll \Lambda, 
\end{equation}
where 
$I_\nu (x)=I_{-\nu}(x)$ are Bessel functions of imaginary argument. 
This result allows us to compare directly with the claims in Refs.~\cite{Ai:2020ptm,Ai:2024cnp}.

The authors of Refs.~\cite{Ai:2020ptm,Ai:2024cnp} observed that {\em i)} when taking $V\to \infty$ after summation over the winding numbers, correlation functions exhibit CP violation that cannot be removed by field redefinitions, whereas {\em ii)} when taking $V\to \infty$ before summation over the winding number sectors, CP violating phenomena are absent. It was furthermore argued that {\em ii)} is the correct approach if topological quantization emerges from the requirement of finite saddles in the action in infinite space-times. 

Clearly, our results are obtained following the procedure in {\em i)}, that is taking the $V\to\infty$ limit at the very end. 
After all, as we have seen, it is even unnecessary to introduce the sum over topological sectors to establish strong CP violation.
Moreover, topological quantization is inherent in the four-torus compactification we envisaged and does not require finite action saddle points in infinite space-times. 

However, our expressions allow us also to investigate what happens if we follow the 
procedure in {\em ii)}, that is 
taking $V\to\infty$ before summing over the topological sectors.  
We find from the asymptotic expansion 
 of the Bessel function for large arguments: 
\begin{equation}
   \sum_{\nu = -\infty}^{\infty}  e^{i\nu \theta} \left( \frac{m}{|m|} \right)^\nu I_\nu ( V \Sigma |m| ) \approx (2\pi V\Sigma |m|)^{-1/2}\,e^{V\Sigma |m|} \sum_{\nu = -\infty}^{\infty}  
      e^{i \nu ( \theta + \arg m )}  
      \,,\ {\rm for\ }\ V\Sigma |m|\gg 1 \,.
\end{equation}
Since $\sum_{\nu = -\infty}^{\infty}  e^{i \nu ( \theta + \arg m )}=\delta (\bar\theta)/(2\pi)$, this appears to corroborate the observation {\em ii)}, namely that $\bar\theta = \theta + \arg(m)=0$ in this ordering of the limits.  
However, in fact what we have demonstrated actually is that the summation over the topological sectors and the $V\to\infty$ limit do not commute.  It is misleading to first take the 
infinite volume limit and then to sum over the topological sectors.\footnote{A similar observation has been made in the context of the dilute instanton gas approximation in Ref.~\cite{Khoze:2025auv}. Note that the latter was precisely the approximation that the authors of Refs.~\cite{Ai:2020ptm,Ai:2024cnp} have relied upon. Further similar observations have been made in 1+1 dimensional toy models~\cite{Albandea:2024fui,Benabou:2025viy}.} 

Our considerations have so far been been based on the assertion that because of the anomaly 
there is no  $\rm U(1)_{R-L}$ pseudo-scalar Goldstone boson in the chiral limit. 
Witten~\cite{Witten:1979vv} challenged this assertion by observing that for “large  $N_c$”, that is for            $N_c\to \infty$ and $g\to 0$,  with $\lambda \equiv g^2 N_c$ fixed, the anomaly turns off, 
\begin{equation}
     \partial_\mu \left(  \bar{q} \,\gamma^\mu \gamma_5 \, q \right) = 2 \bar q\, m\, i\gamma_5 q + 2 \frac{1}{N_c} \frac{\lambda}{32\pi^2} G_{\mu\nu}^a \tilde G^{\mu\nu}_a
\hspace{3ex} \to \hspace{3ex}
 \partial_\mu \left(  \bar{q} \,\gamma^\mu \gamma_5 \, q \right) = 2 \bar q\, m\, i\gamma_5 q .
\end{equation}
As will be reviewed next, 
it is in the sense of the  $1/N_c$ expansion that one can say that the anomaly gives a mass to a pseudo-scalar boson that would have been massless in the chiral limit. 

Within the $1/N_c$ expansion we have to take into account explicitely the angular field $\phi(x)$ whose particle excitation is the $\rm U(1)_{R-L}$ Goldstone boson\footnote{We denote it $\eta^\prime$, in analogy to the corresponding meson in realistic  $N_f=3$ QCD.} arising in the 
large-$N_c$ and chiral limit: $U(x) = e^{-i\phi (x)} \in {\rm U(1)_{R-L}}$.
For large volumes and small quark mass the partition function can be written in this case as~\cite{Leutwyler:1992yt}
\begin{equation}
    Z = \int \left[ dU \right] \exp \left\{ -\int d^4x\, {\mathcal L}_{\rm eff} (U,\partial U,\partial^2 U,\ldots ;me^{i\theta})\right\} ,
\end{equation}
where the effective Lagrangian characterizing the low-energy structure of the theory in the large-$N_c$ limit is of the form~\cite{Rosenzweig:1979ay,Witten:1980sp,DiVecchia:1980yfw}
\begin{equation}
\label{eq:L_eff_large_N_c}
    {\mathcal L}_{\rm eff} = \frac{F^2}{4} \partial_\mu U^\ast \partial_\mu U - \Sigma \,\Re \left( m U^\ast \right) + \frac{\tau}{2} \left( i\ln U - \theta \right)^2 ,
\end{equation}
involving three low-energy parameters: the decay constant, $F \sim N_c^{1/2} \Lambda$, the parameter $\Sigma \sim N_c \Lambda^3$ determining the quark condensate in the chiral limit at infinite volume, and the topological susceptibility $\tau \sim \Lambda^4$ of pure 
${\rm SU}(N_c)$ Yang-Mills theory.\footnote{The effective Lagrangian~\eqref{eq:L_eff_large_N_c} represents an exact result in the following sense~\cite{Leutwyler:1992yt}:
Expand the full effective Lagrangian as a series in derivatives of $U(x)$ and in powers of the quark mass. In addition, expand the effective coupling constants appearing in this series in powers of $1/N_c$. 
This procedure yields contributions of the schematic form 
$\partial^{n_1} m^{n_2} N_c^{1-n_3}$. 
The expansion is organized by first grouping together all terms with a fixed value of 
$n \equiv \frac{1}{2}n_1 + n_2 + n_3 - 1$, and 
only then sum over $n$. Within this counting scheme, all three terms in \eqref{eq:L_eff_large_N_c} are of order $n=0$.} 

For $V\gg m_{\eta^\prime}^{-4}$, the fluctuations in $U(x)$ freeze and 
the partition function reduces to the contribution from the ground state,
\begin{equation}
\label{eq:Z_one_flavor_large_N_c}
Z  = \exp \left\{ - V \min_{U} 
\left\{- \Sigma \,\Re \left( m U^\ast \right) + \frac{\tau}{2} \left( i\ln U - \theta \right)^2 \right\} \right\}
,
\hspace{3ex} {\rm for\hspace{3ex} }\ V  \gg m_{\eta^\prime}^{-4} \ {\rm and\ } |m|\ll \Lambda .
\end{equation}
It is convenient to rewrite the vacuum energy density appearing in \eqref{eq:Z_one_flavor_large_N_c} in terms of the angular field $\phi$: 
\begin{equation}
\label{eq:epsilon_vac_one_flavor_large_N_c}
    \epsilon_{\rm vac} 
     \equiv    
\min_{\phi} 
\left\{ - \Sigma \, |m| \cos (\phi + \arg(m)  ) + \frac{\tau}{2} \left( \phi - \theta \right)^2 
\right\}
    \equiv  \min_{\phi} {\mathcal V_{\rm eff}}(\phi ) 
 .
\end{equation}
We will show now that the relative importance of the first and second term of the effective potential ${\mathcal V_{\rm eff}}(\phi )$ determines the fate of CP violation. 

Let us consider first the case where the second term in $\epsilon_{\rm vac}$ dominates, 
which occurs for $\tau\gg \Sigma |m|$, corresponding to $|m|/\Lambda \ll 1/N_c\ll 1$.
In this case, the potential is minimized at $\phi = \theta$, such that the previous result 
\eqref{eq:vac_energy_one_flavor_chiral_expansion} for the vacuum energy,     
$\epsilon_{\rm vac} 
= - \Sigma\, |m| \cos \bar\theta$, 
as well as the expressions  \eqref{eq:chiral_quark_condensates_one_flavor_rl},   \eqref{eq:chiral_quark_condensates_one_flavor_lr}, \eqref{eq:quark_condensates_one_flavor}, and \eqref{eq:pseudo_scalar_gluon_condensate_one_flavor} for the chiral quark and pseudo-scalar gluon condensates, respectively, are recovered. Moreover, 
introducing the field excitation around this minimum, $\eta^\prime \equiv \phi - \theta$, and using the identity  $\cos ( \eta^\prime + \bar\theta ) = \cos \bar\theta  \cos \eta^\prime  - \sin \bar\theta  \sin \eta^\prime$, we can rewrite the 
effective Lagrangian \eqref{eq:L_eff_large_N_c} as 
\begin{equation}
    {\mathcal L}_{\rm eff} = \frac{F^2}{4} \partial_\mu \eta^\prime \partial_\mu \eta^\prime 
+    \frac{\tau}{2}    \eta^{\prime 2}
- \Sigma \, |m| \left\{ \cos \bar\theta  \cos \eta^\prime  - \sin \bar\theta  \sin \eta^\prime \right\} . 
\end{equation}
The $\eta^\prime$ has a mass of order $m_{\eta^\prime}\sim \sqrt{\tau}/F$, which persists in the chiral limit.  
The term  $\Sigma |m|  \sin \bar\theta  \sin \eta^\prime $   summarizes all CP violating   $\eta^\prime$  amplitudes. We conclude: $N_f=1$ ${\rm SU}(N_c)$ Yang-Mills theory, in the parameter range $|m|/\Lambda \ll {1}/{N_c}\ll 1$, predicts CP violation,  
 as long as $\Sigma |m| \sin \bar\theta \neq 0$.\footnote{In the parameter range $|m|/\Lambda\gg 1$, one may integrate out the quark and consider pure SU($N_c$) gluodynamics with a $\theta$ term, which is expected to have a mass gap $M_0\sim \Lambda$. In the large-$N_c$ limit (and believed more generally), its vacuum energy density, 
$\epsilon_{\rm vac}\equiv -\lim_{V\to \infty}(1/V)\ln Z$, is
a multi-branched function because of many candidate vacuum states that all become stable (but not degenerate) for $N_c\to\infty$: 
$ \epsilon_{\rm vac}(\theta )  =    \min_{k\in \mathbb{Z}} N_c^2 f((\theta + 2\pi k)/N_c)$~\cite{Leutwyler:1992yt,Witten:1998uka}. 
It is approximately quadratic in $\theta$,
$\epsilon_{\rm vac}(\theta ) = N_c^2f(0 ) + \frac{1}{2} \tau \theta^2 + {\mathcal O}(\theta^4/N_c^2)$~\cite{Leutwyler:1992yt}, showing that CP is violated for generic $\theta\neq 0$. 
The multi-branched structure, which is crucial to restore the required $2\pi$ periodicity of the partition function, appears to be beyond the reach of canonical quantization techniques.
This casts strong doubts on the recent claim that there is no CP violation in pure 
gluodynamics~\cite{Ai:2024vfa} because it was based on those techniques. 
}     

Let us consider now the case where the first term in ${\mathcal V}_{\rm eff}(\phi )$ dominates, which occurs for $\tau \ll \Sigma |m| \ll 1$, corresponding to ${1}/{N_c} \ll {|m|}/{\Lambda}  \ll 1$. 
In this case, the potential is minimized at $\phi + \arg (m) = 0 $, such that effective Lagrangian for excitations around the minimum  reads 
\begin{equation}
    {\mathcal L}_{\rm eff} = \frac{F^2}{4} \partial_\mu \eta^\prime \partial_\mu \eta^\prime - \Sigma \, |m| \cos \eta^\prime  
    .
\end{equation}
There is no CP violation! 
The $\eta^\prime$ has a mass $m_{\eta^\prime}\sim \sqrt{\Sigma |m|}/F$ which vanishes in the chiral limit. It is an axion!  
We conclude: $N_f=1$ ${\rm SU}(N_c)$ Yang-Mills theory, in the parameter range ${1}/{N_c} \ll {|m|}/{\Lambda}  \ll 1$, features no CP violation.

At this point we can again comment on Refs.~\cite{Ai:2020ptm,Ai:2024cnp}. We agree with their observation that the low-energy effective Lagrangian admits in principle both possibilities: CP or not CP. However, the question which possibility is realized depends on which values the fundamental parameters have been taken in our real world. 

So far, we have just treated the case $N_f=1$. But the case $N_f>1$ is conceptually only a tiny step away. We just have to include the fields whose particle excitations are the 
$N_f^2 -1$ pseudo-scalar Goldstone bosons from the breaking of ${\rm SU}(N_f)_{\rm R}\times {\rm SU}(N_f)_{\rm L}\times {\rm U}(1)_{\rm R+L}$ to ${\rm SU}(N_f)_{\rm R+L}\times {\rm U}(1)_{\rm R+L}$
into the effective Lagrangian~\cite{Rosenzweig:1979ay,Witten:1980sp,DiVecchia:1980yfw},
\begin{equation}
\label{eq:L_eff_large_N_c_N_f}
    {\mathcal L}_{\rm eff} = \frac{F^2}{4} \partial_\mu U^\dagger \partial_\mu U - \Sigma 
    \,\Re \left( {\rm tr} \left( {\mathcal M} U^\dagger \right) \right) 
    + \frac{\tau}{2} \left( i\ln\det U - \theta \right)^2 ,
\end{equation}
where $U\in {\rm U}(N_f)$. The $\eta^\prime$ is in this case described by $\det U$. One can repeat the steps before to investigate the parameter range where the theory possibly 
violates CP or not, see Refs.~\cite{Rosenzweig:1979ay,Witten:1980sp,DiVecchia:1980yfw,Gaiotto:2017tne}. 

Is real QCD more in the region $m/\Lambda\ll 1/N_c$ or in the region $1/N_c\ll m/\Lambda$? In real QCD $\Lambda\sim 300$\,MeV and $m_u\sim m_d \sim 3$\,MeV, $m_s\sim 100$\,MeV, while $N_c=3$. So, very naively, it appears that we live in a world where $m_i/\Lambda\ll 1/N_c$ and thus CP is violated. This is also corroborated by the fact that 
the $\eta^\prime$ has a mass, $m_{\eta^\prime}\simeq 960$\,MeV, which is indeed of order
$\sqrt{\tau}/F\sim \Lambda$, as expected for the case of strong CP violation.

\section*{Acknowledgments}

Special thanks to J. Jaeckel, V.V. Khoze, and A. Smilga 
for discussions and 
valuable comments on the draft.
This work has been funded by the Deutsche Forschungsgemeinschaft (DFG, German Research Foundation) 
under Germany's Excellence Strategy - EXC 2121 Quantum Universe - 390833306.
This article/publication is based upon work from COST Action COSMICWISPers CA21106, supported by COST (European Cooperation in Science and Technology).


\begin{thebibliography}{99}

\bibitem{Baluni:1978rf}
V.~Baluni,
Phys. Rev. D \textbf{19} (1979), 2227-2230

\bibitem{Crewther:1979pi}
R.~J.~Crewther {\em et al.},
Phys. Lett. B \textbf{88} (1979), 123
[erratum: Phys. Lett. B \textbf{91} (1980), 487]

\bibitem{Abel:2020pzs}
C.~Abel \textit{et al.}, 
Phys. Rev. Lett. \textbf{124} (2020) no.8, 081803
[arXiv:2001.11966 [hep-ex]].

\bibitem{Ai:2020ptm}
W.~Y.~Ai {\em et al.},
Phys. Lett. B \textbf{822} (2021), 136616
[arXiv:2001.07152 [hep-th]].

\bibitem{Ai:2024cnp}
W.~Y.~Ai {\em et al.},
Universe \textbf{10} (2024) no.5, 189
[arXiv:2404.16026 [hep-ph]].

\bibitem{Shifman:1979if}
M.~A.~Shifman, A.~I.~Vainshtein and V.~I.~Zakharov,
Nucl. Phys. B \textbf{166} (1980), 493-506.

\bibitem{Leutwyler:1992yt}
H.~Leutwyler and A.~V.~Smilga,
Phys. Rev. D \textbf{46} (1992), 5607-5632

\bibitem{Crewther:1979rs}
R.~J.~Crewther,
NATO Sci. Ser. B \textbf{55} (1980), 529

\bibitem{tHooft:1981nnx}
G.~'t Hooft,
Commun. Math. Phys. \textbf{81} (1981), 267-275

\bibitem{Khoze:2025auv}
V.~V.~Khoze,
[arXiv:2512.06827 [hep-ph]].

\bibitem{Albandea:2024fui}
D.~Albandea {\em et al.},
Phys. Rev. D \textbf{110} (2024) no.9, 094512
[arXiv:2402.17518 [hep-lat]].

\bibitem{Benabou:2025viy}
J.~N.~Benabou {\em et al.},
[arXiv:2510.18951 [hep-ph]].

\bibitem{Witten:1979vv}
E.~Witten,
Nucl. Phys. B \textbf{156} (1979), 269-283

\bibitem{Rosenzweig:1979ay}
C.~Rosenzweig, J.~Schechter and C.~G.~Trahern,
Phys. Rev. D \textbf{21} (1980), 3388

\bibitem{Witten:1980sp}
E.~Witten,
Annals Phys. \textbf{128} (1980), 363

\bibitem{DiVecchia:1980yfw}
P.~Di Vecchia and G.~Veneziano,
Nucl. Phys. B \textbf{171} (1980), 253-272

\bibitem{Witten:1998uka}
E.~Witten,
Phys. Rev. Lett. \textbf{81} (1998), 2862-2865
[arXiv:hep-th/9807109 [hep-th]].

\bibitem{Ai:2024vfa}
W.~Y.~Ai, B.~Garbrecht and C.~Tamarit,
[arXiv:2403.00747 [hep-th]].

\bibitem{Gaiotto:2017tne}
D.~Gaiotto  {\em et al.},
JHEP \textbf{01} (2018), 110
[arXiv:1708.06806 [hep-th]].

\end{thebibliography}
\end{document}